\documentclass[format=acmtog, review=false, screen, authorversion=false]{acmart}
\usepackage{dblfloatfix}    
\usepackage{tabularx}
\usepackage{booktabs} 
\usepackage{caption}
\usepackage{ragged2e}
\usepackage[ruled]{algorithm2e} 

\SetAlFnt{\small}
\SetAlCapFnt{\small}
\SetAlCapNameFnt{\small}
\SetAlCapHSkip{0pt}
\IncMargin{-\parindent}

\setcopyright{none}      
\renewcommand\footnotetextcopyrightpermission[1]{}
\acmDOI{}
\acmPrice{}
\acmISBN{}
\copyrightyear{}
\acmVolume{}
\acmNumber{}
\acmSubmissionID{}
\acmArticleSeq{0}

\authorsaddresses{Permission to make digital or hard copies of all or part of this work for personal or classroom use is granted without fee provided that copies are not made or distributed for profit or commercial advantage and that copies bear this notice and the full citation on the first page. To copy otherwise, or republish, to post on servers or to redistribute to lists, requires prior specific permission and/or a fee. Copyright \textcopyright JOCSE, a supported publication of the Shodor Education Foundation Inc.}
\begin{document}
\title{\textbf{Automatic Feature Selection in Markov State Models Using Genetic Algorithm}}

\author{Q\lowercase{ihua} C\lowercase{hen}$^{1, a}$, J\lowercase{iangyan} F\lowercase{eng}$^{1, b}$, S\lowercase{hriyaa} M\lowercase{ittal}$^c$}
\author{D\lowercase{iwakar} S\lowercase{hukla}$^{*, b, c, d}$}

\affiliation{\\$^a$Department of Material Science and Engineering, $^b$Department of Chemical and Biomolecular Engineering, $^c$Center for Biophysics and Quantitative Biology, $^d$National Center for Supercomputing Applications, University of Illinois at Urbana-Champaign, Urbana, IL 61801, USA. \\
$^1$ Qihua Chen and Jiangyan Feng contributed equally to this work. $^*$ Correspondence and requests for materials should be addressed to D.S. (email: diwakar.shukla@shuklagroup.org)}

\begin{abstract}
Markov State Models (MSMs) are a powerful framework to reproduce the long-time conformational dynamics of biomolecules using a set of short Molecular Dynamics (MD) simulations. However, precise kinetics predictions of MSMs heavily rely on the features selected to describe the system. Despite the importance of feature selection for large system, determining an optimal set of features remains a difficult unsolved problem. Here, we introduce an automatic approach to optimize feature selection based on genetic algorithms (GA), which adaptively evolves the most fitted solution according to natural selection laws. The power of the GA-based method is illustrated on long atomistic folding simulations of four proteins, varying in length from 28 to 80 residues. Due to the diversity of tested proteins, we expect that our method will be extensible to other proteins and drive MSM building to a more objective protocol. 
\end{abstract}

\maketitle
\pagestyle{plain}
\thispagestyle{plain}
\textbf{\section{Introduction}}
Molecular Dynamics (MD) simulation, first introduced by Alder and Wainwright\cite{alder1957phase} in the late 1950's, has evolved into a major technique to study the detailed actions and mechanisms of proteins\cite{lane2013milliseconds, plattner2017complete, lindorff2011fast, doerr2016htmd, moffett2018using}. Based on Newton's equations of motion, MD simulations can describe protein dynamics in unprecedented spatial and temporal resolution. However, one of the major challenges for MD simulations are the analysis of high dimensional data and the incompatibility between timescales accessible to MD simulation and that are functionally relevant\cite{shukla2016conformational, vanatta2015network, shukla2014activation, kohlhoff2014cloud, lawrenz2015cloud}. Markov State Models (MSMs)\cite{husic2018markov, shukla2015markov, pande2010everything} have recently been used to address the aforementioned issues by predicting protein dynamics at long timescales from a pool of short MD simulations. The MSM itself is a "transition probability matrix"\cite{bowman2014introduction}, describing mathematically the memoryless transitions between metastable states. To construct a MSM, raw MD trajectories are first transformed from their Cartesian coordinates to features, such as dihedral angles\cite{husic2016optimized, mittal2018recruiting} or pairwise contact distances of a protein. This step is often called "featurization". The dimensionality of these features may be further reduced through dimensionality reduction step. One commonly used method is time-structure independent components analysis (tICA), which creates linear combinations of input features by maximizing their decorrelation time\cite{schwantes2013improvements, perez2013identification, m2017tica, lapidus2014complex, schwantes2016markov}. With a properly constructed MSM, useful thermodynamic and kinetic properties of the dynamic process can be extracted. Despite the attractive feature of MSMs, the thermodynamics and kinetics predicted by MSMs are highly sensitive to which features are selected to discretize the configuration space\cite{doerr2016htmd, beauchamp2011msmbuilder2, mcgibbon2015mdtraj}. Ideally, features should be chosen to capture the slowest motions of the protein, which are usually the most interesting or important processes. However, determining an optimal set of features remains a considerable challenge especially when a protein system is sufficiently complex. Here, we show that machine 

Currently, there are two major ways of selecting features in terms of "contact featurization", where pairwise contact distances of a protein are used as features. One is using all pairwise contact distances of a protein as features. In principle, no important information about the system is missed out since all the contact distances are considered. However, it is costly to calculate all distances even for a small protein. For a protein system with R residues, the total number of distances among each other will be $R$($R$-1)/2, which creates a heavy load of calculation on computers. In addition, irrelevant features that do not contribute to the dynamics process may lead to the poor generalization performance of the model. Thus, using all available features may degrade the performance of the MSM both in speed (due to high dimensionality) and accuracy (due to irrelevant information). Alternatively, the most commonly used method is choosing a subset of contact distances based on human intuition. Consequently, the thermodynamics and kinetics extracted from MSMs can be biased by the manually chosen features. In summary, either way is appropriate for the selection of features and a more convenient, accurate and automated method for feature selection is necessary. A variety of machine learning methods have been recently reported for dimensionality reduction and/or feature selection for molecular dynamics datasets\cite{mittal2018recruiting}. However, the use of these ideas for automatic feature selection in building MSMs has not been explored.

Here, we present a genetic-algorithm based method to select an optimal set of residue pair distances for contact featurization. Genetic algorithm (GA) is one of the advanced methods to help with dealing feature selection problems in data science. First proposed by John H. Holland\cite{goldberg1988genetic, holland1992adaptation}, GA is a heuristic and adaptive simulation algorithm that evolves the most fitted solution to a problem based on Darwinian natural selection laws. GA has been broadly applied to help with function optimization\cite{taherdangkoo2013efficient}, protein folding prediction\cite{kenneth2012protein, unger1993genetic}, multiple sequence alignment\cite{gondro2007simple} and more scientific investigations\cite{hoque2017genetic}. In nature, useful traits in genes tend to be preserved in offspring for a higher survival probability. Like the real cases in nature, better solutions to a problem can be derived by GA according to this principal. In our case, each gene represents the alpha carbon distance between a residue pair, and chromosomes are combinations of residue pair distances. To seed the whole process, we randomly select one residue pair distance as the starting point of the GA. The adaptability of each chromosome (a set of residue pair distances) is quantitatively expressed as fitness scores in GA. In this study, we use generalized matrix Rayleigh quotient (GMRQ) score as the fitness score. GMRQ was recently introduced to quantitatively evaluate MSMs based on its distance from a theoretical upper limit\cite{husic2017note, mcgibbon2015variational, noe2013variational}. The higher the GMRQ score is, the more prominent the MSM is to capture the slow underlying dynamical motions while a low GMRQ score indicates that the MSM is not able to reveal the slow dynamics of the system. Therefore, the goal of our method is optimizing a set of residue pairs that gives the highest GMRQ score. The framework of our GA-based method is adapted from the "Optimal Probes" method proposed by Mittal and Shukla\cite{mittal2017predicting}. In their study, an optimal choice of residue pairs, capturing the slow conformational dynamics, is successfully predicted for double electron-election resonance spectroscopy, an experimental technique capable of detecting conformational changes by monitoring the distance between electron spins.

In this method, we (1) perform contact featurization for each set of residue pair alpha carbon distances, (2) use tICA to further reduce the dimensionality of the data, (3) construct MSMs based on the reduced dimensionality, and (4) calculate GMRQ for each set of residue pair distances to evaluate the MSMs. Based on the GMRQ score, the combination of residue pair distances will be updated. The algorithm will then go back to step (1) to repeat the whole process until reaching user specified number of iterations. In the end, the set of distances with the maximum GMRQ score is chosen as an optimal set of residues for the construction of the "best MSM". To evaluate of our method, we test the GA-based method on four folding proteins with the size ranging from 28 residues to 80 residues. Our experimental results show that the method yields comparable and even better accuracy compared with using all available features. To our knowledge, this is the first attempt to automatically select proper MSM features for analysis. The GA-based method described here to larger proteins undergoing conformational changes can be extended.
\textbf{\section{Theory and Methods}}
\par\noindent\textbf{Molecular Dynamics (MD) Simulation Dataset.} MD simulation datasets of the four folding proteins for analysis were generated by Lindorff-Larsen \textit{et al}\cite{lindorff2011fast}. The four proteins (BBA, Villin, WW domain and $\lambda$-repressor) vary in length from 28 to 80 amino acids. More details of the simulations are summarized in Table 1. For the analysis, we retain all the trajectory frames. Three small  proteins (BBA, Villin and WW domain) are chosen to evaluate the proposed method and the best GMRQ achieved using all contact distances serves as the benchmark. The 80-amino-acid $\lambda$-repressor is used to test the feasibility of the method on large proteins, as using all distances is impractical.
\begin{table}[h]
\caption{Protein and trajectory information.}
\begin{tabular}{ l  c  c c }
\hline
\hline			
Protein & PDB & Residues & Total simulation time ($\mu$s) \\
\hline  
BBA & 1FME & 28 & 325 \\ 
Villin & 2F4K & 47 & 429 \\
WW domain & 2F21 & 35 & 1137\\
$\lambda$-repressor & 1LMB & 80 & 643\\
\hline  
\hline  
\end{tabular}
\end{table}
\\
\textbf{Markov State Models (MSMs).} In this study, the goal is optimizing a set of residue pair distances to build the best MSM based GMRQ. MSMs are kinetic models that reveal the dynamics of a system\cite{hummer2014optimal, prinz2011markov, plattner2017complete, pande2010everything, bowman2014introduction}. An MSM describes a network of metastable conformational states and reveals the probabilities of each state performing jumps from one to another over an appropriate time resolution ($\tau$, also called lag time). The jumps are memoryless, which means the probability to transit to the present state is not dependent on the previous ones. Such information is presented in a "transition probability matrix" by MSM, where an $n\times n$ square matrix depicts the transitions among $n$ states\cite{bowman2014introduction}. The probability of each jump can be expressed according to the equation below: 
\begin{equation}
p_j(t+\tau)=\sum_{i=1}^{n} p_i(t)T_{ij}(\tau)
\end{equation}
The equation can also be expressed in a matrix form: 
\begin{equation}
p^T(t+\tau)=p^T(t)T(\tau)
\end{equation}
where $p_i(t)$ is a population vector whose elements show the probability at time $t$, $p_j(t+\tau)$ is a population vector after time $\tau$, $T_{ij}(\tau)$ is the probability to jump from state $i$ to state $j$ and $T(\tau)$ is the transition probability matrix that $T(\tau)\in \mathbb{R}^{n\times n}$. Further details of the transition matrix can be found in literatures\cite{bowman2014introduction, shukla2015markov}.

The transition probability matrix  can be decomposed into eigenfunctions and eigenvalues shown below: 
\begin{equation}
T(\tau)\circ\psi_i=\lambda_i\psi_i
\end{equation}
where $\psi_i$ is the eigenfunction and $\lambda_i$ are the real eigenvalues that $\lambda_i\leq1$, arranged in descending order. 

Here, each step of the MSM building process used in this study is described in detail. All the hyperparameters (e.g. the number of tICA components, tICA lagtime, the number of clusters, the number of MSM timescales and MSM lagtime) are shown in Table 2.
\begin{enumerate}
  \item Featurization. To construct an MSM, the first step is to process the datasets that we plan to work on. In our case, we use the MD simulation data sets listed in Table 1. The datasets are given in the form of MD trajectories, which present series of motion of the protein atoms in a frame-wise arrangement. Because the simulated movements recorded in Cartesian coordinates are not ideal for analysis, and too much noise not relevant to our study may be included, it is better to interpret the data in other ways. As a result, a lot of reasonable metrics such as dihedral angle\cite{husic2016optimized} and contact distances between residue pairs are used to featurize the data. The featurization method we choose here is contact distance analysis. By using such technique, more useful information can be extracted from the redundant MD trajectories. Again, our goal of this study is to optimize the choice of residue pairs for contact distance calculation, so that an MSM with more information and less noise can be found by this method. The method outlined in this study could be applied to any chosen set of features calculated using simulation data. 
  \item Dimensionality reduction. We further processed our featurized data by tICA so as to reduce the dimensionality of the data. After featurization, the featurized data were projected onto linear subspaces of the slowest dynamics. The components of tICA are termed time structure-based independent components (tICs), which are linear combination of the input features (a set of contact distances in our case). Top tICs capture the slowest motion captured by tICA and usually represent the most interesting dynamics\cite{schwantes2013improvements, perez2013identification, m2017tica, lapidus2014complex, schwantes2016markov}.
  \item Clustering. We perform mini-batch k-means clustering on the processed data. Clustering refers to the coarse graining analysis that groups certain datasets based on their similarities, so that macrostates can be formed to be better understood. Commonly used clustering algorithms, such as mini-batch k-means\cite{bowman2014introduction, mcgibbon2015variational, moffett2017allosteric}, mini-batch k-medoids\cite{chodera2007automatic, friedman2001elements} and k-centers\cite{beauchamp2011quantitative, lapidus2014complex}, have shown similar performance when the data is preprocessed with tICA\cite{schwantes2013improvements, perez2013identification, m2017tica, mittal2018recruiting}.
  \item MSM construction. After the clustering, a MSM can be built based on the processed datasets. The process was implemented in a Python environment and the software involved to produce the analysis above include Numpy\cite{ascher2001numerical}, MDTraj\cite{mcgibbon2015mdtraj} and MSMBuilder3.8\cite{beauchamp2011msmbuilder2}. 
\end{enumerate}
\,
\textbf{Generalized Rayleigh Quotient (GMRQ).} In short, an ideal MSM should successfully identify the slowest dynamics of the protein. Because the state decomposition mentioned above reveals the dynamical processes in the system, the identification of true eigenfunction and eigenvalues become the major problem for scientists to solve. A more quantitative method is needed to help evaluate and find the true state decomposition, which is directly related to the choice of metrics in the featurization stage. 

To help solve this problem, GMRQ was introduced as a quantitative way of evaluating the quality of an MSM\cite{mcgibbon2015variational, noe2013variational, mcgibbon2015variational}. GMRQ is derived from the variational principle that adds up the first $m$ eigenvalues, which denote the slowest $m$ dynamical processes in the system. The variational principles set an upper boundary\cite{husic2017note, mcgibbon2015variational, noe2013variational} for the total sum of real eigenvalues shown below:
\begin{equation}
GMRQ\equiv \sum_{i=1}^m \hat{\lambda}_i \leq \sum_{i=1}^m \lambda_i
\end{equation}
where the $\hat{\lambda}_i $ is the estimated eigenvalue and the $\lambda_i$ is the real eigenvalue. In this study, as we try to maximize GMRQ score to approach the upper boundary, the larger the GMRQ score we get, the closer we are to the slowest dynamics of the protein.

To help avoid overfitting, cross-validation must be applied to evaluate our GMRQ scores. The dataset from the MD simulation is split into a training set and a test set. The training set is first used to estimate the model parameters such as the eigenvalues, then the estimated model is applied to score its performance in the test set. In this way, the model will not be biased by overfitting the data onto the model. The process of deriving GMRQ scores is achieved by Osprey package\cite{mcgibbon2016osprey} and the recruited parameters are shown in Table 2. Mean GMRQ of five cross-validation iterations are used for the analysis.\\
\\
\begin{figure}[!h]
\includegraphics[width=0.8\linewidth]{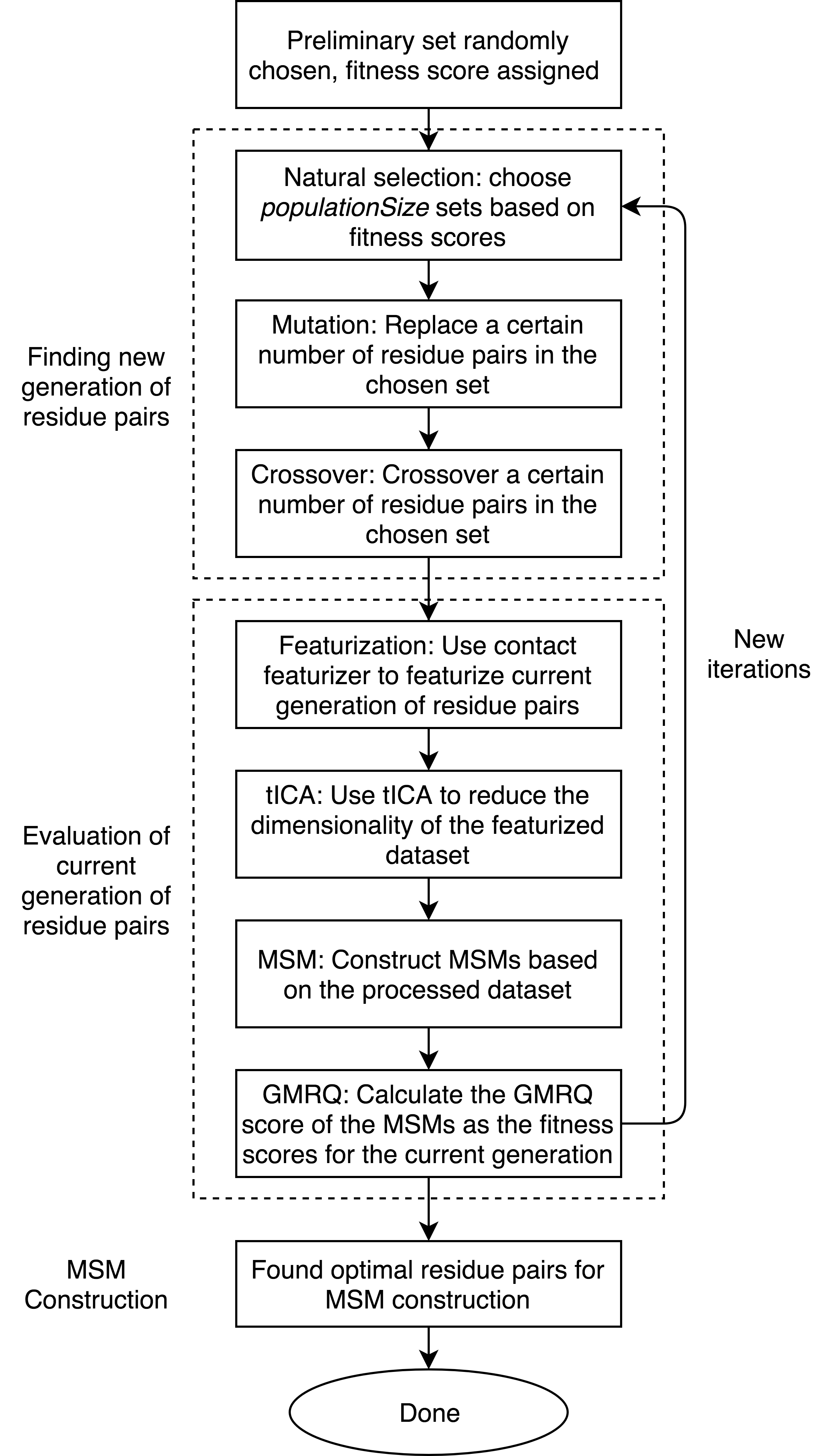}
\caption{The flow chart showing the whole process of our GA-based method. }
\end{figure}
\\
\textbf{Genetic-algorithm-based Method for automatic feature selection in Markov State Models (MSMs).} To simulate the natural selection process according to Darwinian laws, we must decide how the natural selection principles are implemented in our algorithm. In this section, we introduce our basic operators of GA, the framework that we follow to perform GA, and the protocol we adopt to finally generate optimal residue pairs. The construction of the GA is based on the work of Mittal and Shukla\cite{mittal2017predicting}.

In the field of programming, operators refer to the actions to take during each step of execution of the algorithm. The basic operators in our study are composed of natural selection, mutation, and crossover. In the following section, we discuss our method to help predict an optimized set of residue pair distances for MSM construction using genetic algorithm. We also provide the series of steps as a flow chart shown in Figure 1. Some important parameters that are involved in these steps are: \textit{populationSize}, \textit{percentMutation} and \textit{percentCrossover}. These parameters can be changed according to user's need. 
\begin{enumerate}
\item	A set of all possible residue pairs is identified. $R$($R$-1)/2 residue pairs for a protein with $R$ residues.
\item	\textit{populationSize} preliminary sets of residue pair are randomly selected from the set of all possible residue pairs for the first iteration. Each set contains only one residue pair as the starting point for selection. These sets of residue pairs serve as the initial generation $G_0$ and are assigned fitness scores of 0.
\item	Natural selection is performed to choose new generation of residue pairs according to their fitness scores. The natural selection operator corresponds to the reproductive process in nature, which selects genomes with ideal traits for breeding offspring. In our case, we define a parameter \textit{populationSize} that describes the number of elements randomly chosen from the parental set for a new generation $G_{new}$. 
\item	Mutation is performed to maintain diversity to the current generation of residue pair selections. The mutation operator corresponds to the mutation process in nature to increase genetic diversity. In our version of GA, we define a parameter \textit{percentMutation} to maintain a ratio of mutation in our combination of residue pairs. During the mutation step, the number of residue pairs to be mutated are generated by (\textit{percentMutation} $\times$ \textit{populationSize})$/100$ from $G_{new}$ and those residue pairs are randomly replaced by other residue pairs that are excluded in the $G_{new}$. 
\item	Crossover is performed to add more diversity to the current generation. The crossover operator corresponds to the natural recombination process of chromosomes. Here, we define another parameter \textit{percentCrossover} as the percentage of crossover in our combination of residue pairs. The number of residue pairs to perform crossover is generated by (\textit{percentCrossover} $\times$ \textit{populationSize})$/100$ from $G_{new}$. The residue pair distance sets will then be swapped according to the number calculated before to create a new combination of residue pair distances.  
\item	Evaluations are performed to assign fitness scores to the newly generated residue pairs. MSMs are constructed based on contact featurization using the current generation of residue pairs, and GMRQ scores are calculated accordingly to serve as  fitness scores. 
\item	If more iterations are designed to be finished, the next iteration should restart at step (3) and use the current generation of residue pairs as $G_0$. As the iteration number increases, the fitness scores for the selection of residue pairs should show a convergence of fitness scores.  
\end{enumerate}

All the parameters used in this study are organized in Table 2.

\begin{table}[h]
\caption{Model hyperparameteres.}
\begin{tabularx}{\linewidth}{@{\extracolsep{\fill} } l  c  c }
\hline\hline			
\textbf{Featurization} &  & \\
$\alpha$-carbon contact distances &  &   \\ \hline
\textbf{Decomposition} & Components & Lag time (ns)  \\
tICA & 5 & 0.2\\ 
\hline
\textbf{Clustering} & Clusters &   \\ 
Mini-batch k-means & 200 &   \\
\hline
\textbf{Model fitting} & N\_timescales & Lag time (ns)\\
MSM & 5 & 50  \\ 
\hline
\textbf{Scoring} &  &   \\
GMRQ &  & \\
\hline
\textbf{Cross-validation} & Iterations & Test set size  \\
Shuffle \& Split & 5 & 0.5\\
\hline  
\textbf{Genetic algorithm} &  &  \\
Iterations & 40 & \\
populationSize & 20\% & \\
percentMutation & 50\% & \\
percentCrossover & 20\% & \\
\hline
\hline  
\end{tabularx}
\end{table}
\begin{table*}[ht]
\centering
\caption{Comparison of the best GMRQ scores generated and benchmark GMRQ scores from all contact featurization. The fraction of all residue pairs is the fraction of chosen residue pairs in all residue pairs. The best GMRQ refers to the highest GMRQ score that we obtain from MSMs using residue pair distance features given by our genetic algorithm approach, and the benchmark GMRQ score is the GMRQ provided by the MSM constructed with all contact featurization. The deviation column if the deviation of our best GMRQ score from the benchmark GMRQ score. }
\centering
\begin{tabularx}{\linewidth}{ @{\extracolsep{\fill} }lcccccc }
\hline\hline			
 Protein&  Residues & Number of chosen &Fraction of & Best & Benchmark & Deviation (\%) \\
 &   &  distances &pairs (\%) &GMRQ & GMRQ & \\
\hline
BBA (1FME) & 28 & 47 &12.43 &4.445 & 4.239 & +4.80\\ 
Villin (2F4K) & 35 & 61 &10.25 & 3.203 & 3.705 &-13.5\\
WW domain (2F21) & 35& 4 &0.67 & 4.198 & 4.111& +2.12\\
$\lambda$-repressor (1LMB) & 80 & 60 & 1.90 & 4.956 & N/A&N/A \\
\hline\hline  
\end{tabularx}
\end{table*}
\textbf{\section{Results}}
\noindent In this section, we discuss the optimized set of residue pair distances obtained from our GA-based approach. As described in the method part, the unbiased and extensive MD simulation data (>100$\mu s$) simulating the folding process of the proteins is taken from literature\cite{lindorff2011fast}. Preliminary sets of residue pair distance are randomly selected from the set of all the possible residue pairs as the starting point of the genetic algorithms. These sets go through selection, mutation and crossover steps to provide a new generation of residue pair distances. In the setting of GA, we choose a population size of 10\%, mutation percentage of 50\% and crossover percentage of 20\%. Next, the newly generated residue pair distances are used to build MSMs and assign new GMRQ scores (fitness scores) for evaluation. The next iteration will then go back to the selection step and select according to the newly assigned fitness scores. As the process goes through more iterations, the GMRQ scores will converge and a best GMRQ score can be found. 
\begin{figure*}[ht]
\includegraphics[width=0.8\linewidth]{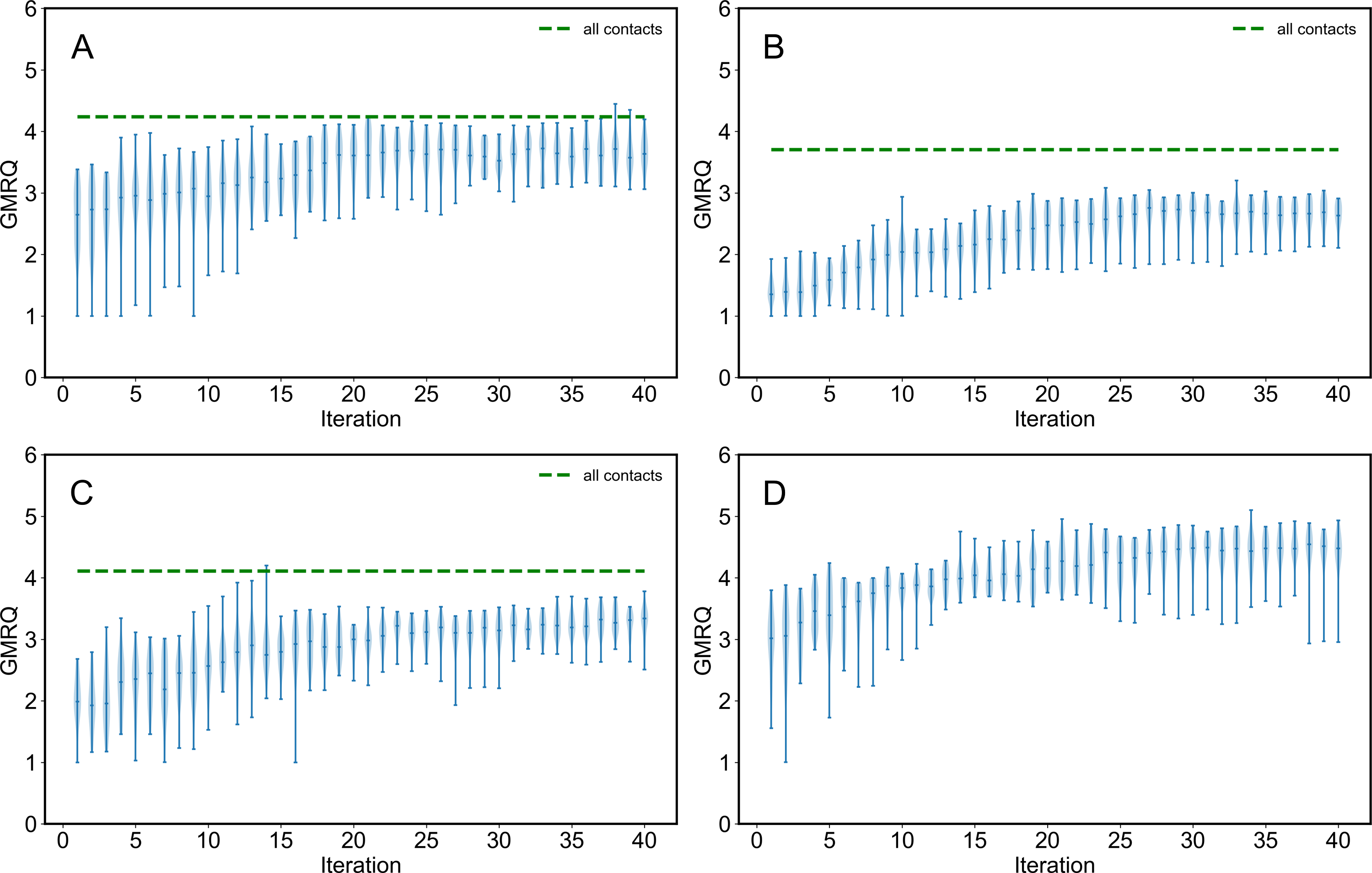}
\caption{GMRQ scores reflecting the MSMs based on the GA-predicted residue pairs. (A) BBA (PDB ID: 1FME), (B) Villin (PDB ID: 2F4K), (C) WW domain (PDB ID: 2F21), (D) $\lambda$-repressor (PDB ID: 1LMB). Green, dashed lines indicate the best GMRQ score corresponding to MSMs based on all contact featurization. Each violin plot shows the increase of GMRQ scores over 40 iterations. In each set of data, the center dot shows the mean values and the vertical line shows the range of this GMRQ data set. }
\end{figure*}
\begin{figure*}[!hb]
\includegraphics[width=\linewidth]{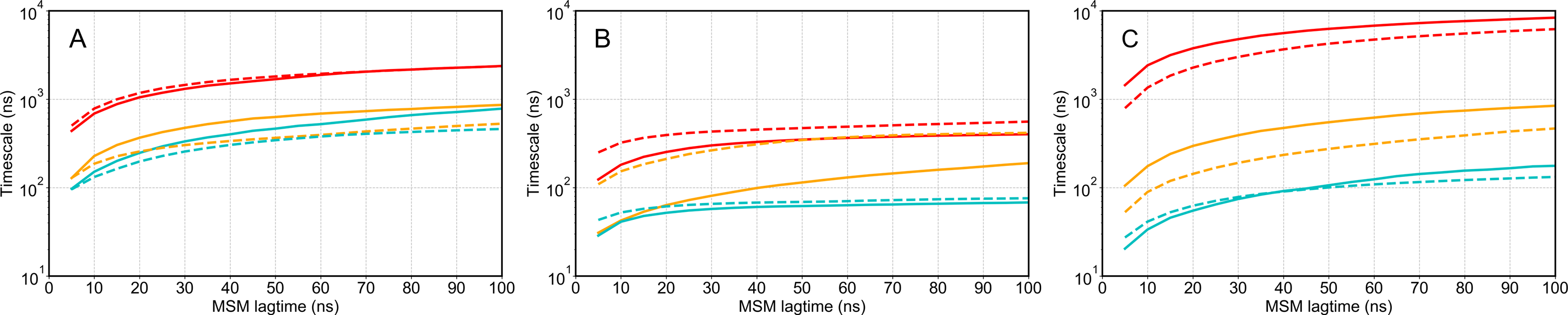}
\caption{The first three slowest implied timescales as a function of MSM lag time. (A) BBA (PDB ID: 1FME), (B) Villin (PDB ID: 2F4K), (C) WW domain (PDB ID: 2F21). The red, yellow and cyan colored lines indicate the slowest, second slowest and third slowest implied timescales, respectively. Dashed lines correspond to the reference value given by the MSMs built on all contacts featurization. Solid lines correspond to the implied timescales given by the MSMs achieved by using the set of distances optimally chosen by our GA-based method.}
\end{figure*}

This method is applied to 4 proteins for demonstration of its functionality: BBA, Villin, WW domain and $\lambda$-repressor. Among the proteins, 3 proteins (BBA, Villin and WW domain) are small proteins, each of which has a residue number that smaller than 40 ($R$ < 40). To examine the effectiveness of our method, we compare the GMRQ scores and implied timescales with their corresponding benchmark values (using all contact distances as features). In the end, we show the ability of our GA-based method to process larger proteins such as $\lambda$-repressor, a protein with 80 residues, which cannot be featurized using all contact distances.
\textbf{\subsection{Our GA-based method proved effectiveness in generating GMRQ scores that are close to the highest possible values given by all contacts featurizer.}}
\noindent We featurize the small proteins (BBA, Villin, WW domain) using all contacts featurization to produce benchmark GMRQ scores for comparison. Benchmark GMRQ scores will serve as a comparable reference to evaluate the performance of our method of using GA to generate optimal residue pairs as featurization metrics. By comparing the best GMRQ scores from our GA-based method to the benchmark GMRQ scores, we are able to check whether our method successfully provides the residue pair sets that depict the slowest process of the protein dynamics. We also apply this method to $\lambda$-repressor, a medium sized protein with 80 residues, to show its ability to process larger proteins. The GMRQ scores are calculated by adding up the eigenvalues of the transition probability matrix provided by MSMs\cite{mcgibbon2015variational, noe2013variational}. The theoretical upper limit of GMRQ score is 6 in all cases\cite{husic2017note, mcgibbon2015variational, noe2013variational}, due to the fact that the number of MSM timescales is chosen to be 5 in the MSM settings. Therefore, in our case, high GMRQ score that approaches 6 usually suggests a better ability of an MSM to capture the slowest process, whereas low GMRQ score implies ineffective state decomposition during the MSM construction process. All information regarding the GMRQ scores and residue pair selection is summarized in Table 3. As shown in Figure 2, all GMRQ scores converged over 40 iterations. In Figure 2A, the highest GMRQ score for BBA is around 4.445, which is higher than the benchmark GMRQ score (4.239). Similar traits are shown by WW domain in Figure 2C that the best GMRQ from GA (4.198) is higher than the benchmark (4.111). However, one exception happens in Villin, shown in Figure 2B. In Figure 2B, the best predicted GMRQ (3.203) does not reach the benchmark (3.705). More iterations for Villin are needed to reach a best GMRQ score that is higher than the benchmark, but there exists a trade-off between the accuracy and computational resource needed.
Overall, the percent variances between our predicted GMRQ score and the benchmark GMRQ score are +4.8\% for BBA, -13.5\% for Villin and +2.12\% for WW domain, in which Villin has the highest difference compared to the other two proteins. 

Similar analysis is applied to $\lambda$-repressor, except that $\lambda$-repressor lacks a benchmark GMRQ score due to its higher number of residues. Hence, there is no reference value to compare in this case. The best GMRQ score given over 40 iterations is around 4.956. Considering that the upper limit of the GMRQ score in this system is 6, we believe that a score of 4.956 is a relatively high GMRQ that effectively captures the slow dynamics of the protein folding mechanisms. Therefore, we can conclude that the method proves its ability to provide the optimal selection of residue pairs for the construction of the best MSM.
\textbf{\subsection{Implied timescale plots show that predicted optimal sets of residue pair distances are able to successfully capture the slowest dynamics in the proteins.} }
\noindent By plotting lag time dependent implied timescale plots, we can quantitatively visualize the slow modes of protein dynamics. Figure 3 shows the comparison between the converged slowest implied timescales provided by all contact featurization and our GA-based method. Again, the reference values are provided by utilizing all residue pair distances as features. Since $\lambda$-repressor is too big for all contacts featurization, there is no benchmark data available and its implied timescale is not shown. In Figure 3A, the slowest implied timescales (solid and dashed red lines) of BBA nearly overlap with each other, indicating that our method has chosen a set residue pair distances that captures the slowest process. In addition, the predicted second and third slowest implied timescales (yellow and cyan) are slower than the corresponding timescale for the benchmarks. In Figure 3B, the predicted implied timescales of Villin has a larger deviation. This inconsistency will be explained and justified in the next paragraph. In the case of WW domain (Figure 3C), we capture a slower timescale than the benchmarks. We find that inclusion of all residue pair distances can add noises to the model, and our GA-based method helps improve the MSM construction by excluding those irrelevant features. 
\begin{figure*}[t]
\includegraphics[width=0.8\linewidth]{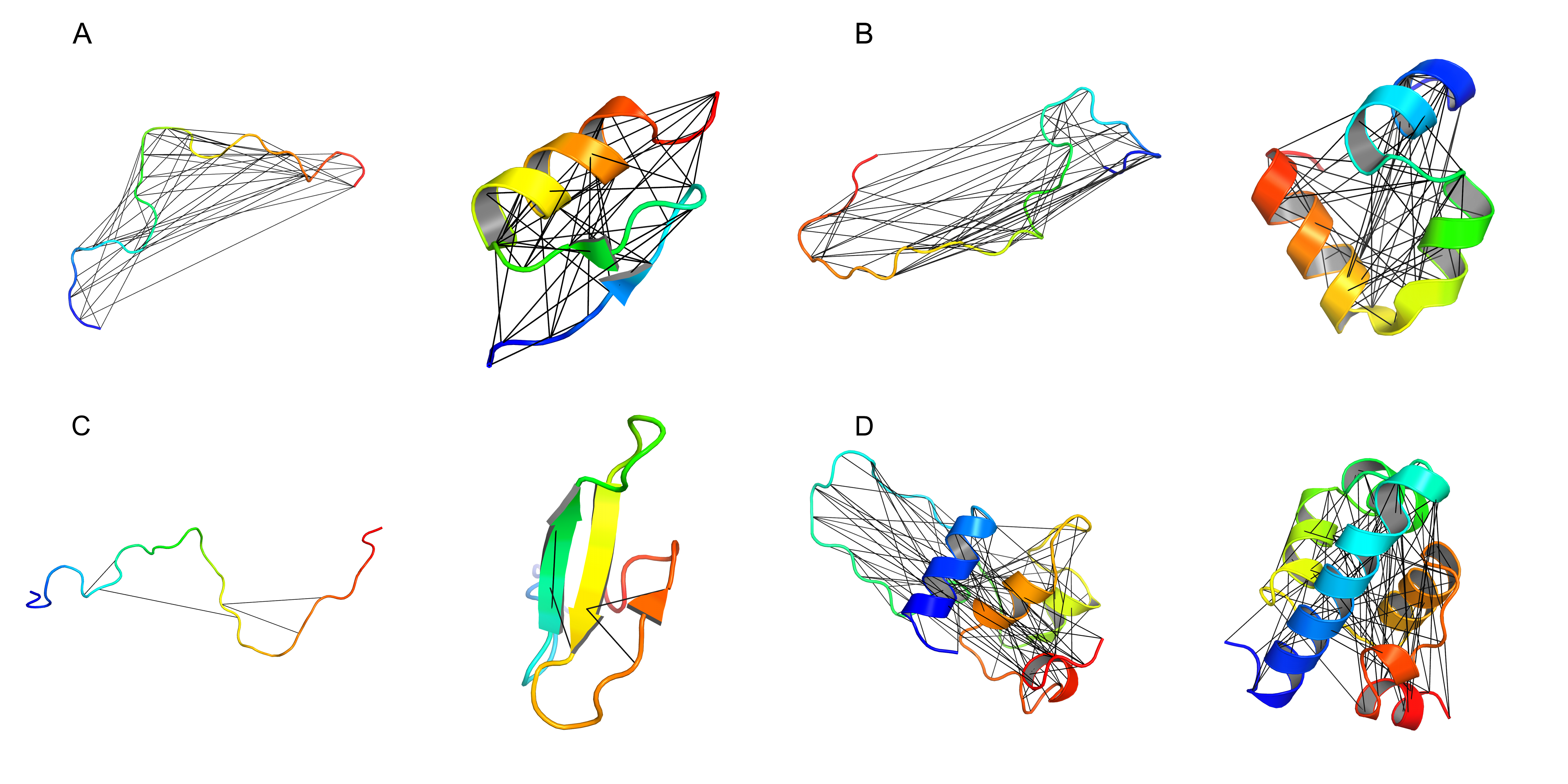}
\caption{GA-chosen residue pairs visualized on the unfolded MD structures and folded crystal structures. (A) BBA (PDB ID: 1FME), (B) Villin (PDB ID: 2F4K), (C) WW domain (PDB ID: 2F21), (D) $\lambda$-repressor (PDB ID: 1LMB). The black lines specify the distances between residue pairs chosen by our GA-based method, which capture the slowest dynamics of the proteins.}
\end{figure*}
\textbf{\subsection{Number of selected distances may reflect the degree of complexity of the protein folding mechanism.} }
\noindent Other than the GMRQ scores and implied timescale plots, more information can be obtained from the sets of residue pair distances. In Table 2, we collect and summarize the number of distances selected by GA and the actual residue numbers in each protein. One interesting thing is that the number of residues in a protein is not necessary correlated to the number of distances needed to capture its slowest dynamics. For example, it can be observed in Table 2 that although both Villin and WW domain have 35 residues in their sequences, WW domain only needs 4 distances of residue pairs while Villin requires 61 distances. This may be due to the complex folding mechanism of Villin. Though both proteins are fast-folding proteins with small numbers of residues, the secondary structure elements in Villin fold more independently without much interactions\cite{mcknight1996thermostable}. Such minimized interaction or minimal frustration makes the folding kinetics fast for Villin, according to the folding funnel theory\cite{bryngelson1995funnels}. Consequently, because the protein folds quickly, this phenomenon suggests a continuous reduction in energy in the folding funnel\cite{bryngelson1995funnels, dill1997levinthal}, which implies multiple parallel pathways during the folding hypothesis\cite{zhu2011evidence}. On the other hand, WW domain folds much slower than Villin\cite{bowman2014introduction} and has more consistent folding pathways\cite{a2012dominant}. The independent features in Villin make it hard for GA to fully capture its slowest dynamics. 

In a previous study, Feng and Shukla\cite{feng2018characterizing} utilized evolution couplings (ECs) as functional features to capture protein folding and conformational dynamics, which gives the similar results for Villin and WW domain. Their work identified that Villin needs 73 ECs and WW domain only needs 5 ECs to fully describe the protein dynamics. They stated that more ECs are needed if the ECs has low correlation. Here, our results show the same trait that Villin requires more features for identification of its slowest dynamic processes, which is reasonable due to the folding complexity of Villin comparing to other fast-folding small proteins. To fully capture the slow dynamics of proteins like Villin, a large number of features should be included from its whole dataset. This is a different scenario comparing to capturing the dynamics of the proteins that needs small numbers of features, which is a problem easier for GA to solve. For proteins like Villin, other methods needs to be explored for a more efficient way to capture the slowest dynamics. Although our method results in some degrees of deviations from the benchmarks (shown in Figure 2B and 3B), it still shows effectiveness in dealing with proteins with complicated kinetics. 

To present our predicted results in a more understandable way, we visualize the optimal sets of residue pairs for all four proteins in Figure 4. Each section (A, B, C and D) of Figure 4 consists of two parts, representing the unfolded and folded structure of the protein respectively. It is easy to notice that the residue pair distances chosen by our method spread out in the protein to capture the complex dynamics of protein folding.

\textbf{\section{Conclusion}}
\noindent Feature selection of MSM construction determines the accuracy of predicted kinetics properties. Currently, the selection of features is done using trial and error. The utilization of GMRQ score enables a quantitative description of the accuracy of MSMs in representing the molecular dynamics observed in a simulation dataset. Using GMRQ score as fitness score, we introduce a GA-based method in order to optimize a set of residue pair distances that produce superior MSMs. In this study, we have shown that our method can provide an automatic, efficient and accurate way to choose the optimal residue pair distances as features for MSMs construction. This significantly improves the efficiency in the overall process of building MSMs while still guarantees the quality of MSMs to capture the slowest protein dynamics. Due to the diversity of tested proteins, our method can be widely applied to other proteins to help with the feature selection process and we anticipate that this method will shift MSM building one step closer to a systemic and objective protocol. It is important to be aware that the underlying assumption of this approach is that the slowest dynamic processes correspond to the process of interest. However, this assumption can be challenged in the case of insufficient sampling or inaccurate force field.

However, the method also has some limitations. The proposed method belongs to the class of wrapper methods for feature selection that find the ``optimal" feature subset by iteratively selecting features based on the classifier performance. The performance of these methods drops significantly for datasets with large number of important but uncorrelated features. Our method also does not perform well on systems with complex dynamics that requires a large number of features to capture the underlying dynamics. In other words, the effectiveness partially depends on the complexity of the conformational changes in the protein, which is shown in the discussion of Villin. As the folding complexity increases, more pathways are available for the protein, so the selection of residue pairs may not fully depict the slowest dynamics of the protein. However, a large number of biologically relevant dynamic processes have been shown to involve only a few important features\cite{feng2018characterizing, shamsi2017enhanced, sultan2014automatic, shukla2016conformational, vanatta2015network, shukla2014activation, kohlhoff2014cloud, lawrenz2015cloud}. In addition, sequence information and crystal structure of the protein should be known, and sufficient amount of MD simulation data should be generated to apply our method. In conclusion, the proposed algorithm, can help identify essential residue pair distances for featurization and exclude noises for MSM construction with high efficiency.

\textbf{\section{Reflection}}
\noindent The year-long Blue Waters Internship enriched my experience in many aspects. This opportunity was rare and precious, especially that I can utilize one of the leading-edge petascale computational resources on the Blue Waters Supercomputer. I was excited to be offered the opportunity to meet other interns to study and practice computational skills together. Starting last summer, I have been involved in a variety of activities, including a two-week educational workshop at University of Illinois at Urbana-Champaign, regular webinars, monthly reports and preparing a manuscript. Majoring in Material Science, I joined the internship with limited computational experience. However, I quickly gained essential skills and became adept with the help of internship coordinators, my research advisor and mentors in the lab. In addition, working on the projects helped me to be familiar with the life in a research group and be better prepared for the graduate school. My presentation skills were improved through attending group meetings and poster sessions. I also practiced my writing skills through regular progress reports and writing this manuscript. Overall, the past year was a busy year, but it has became a unique experience in my undergraduate studies. 
\\
\begin{acks}
The authors thank D. E. Shaw Research for MD simulation trajectories (BBA, Villin, WW domain and $\lambda$-repressor). Q.C. would like to thank the support of the Shodor Education Foundation and the Blue Waters Student Internship Program. J.F. would like to thank Chia-chen Chu fellowship for support. S.M. is supported by the CSE Fellows Program, Computational Science and Engineering at University of Illinois, Urbana-Champaign, Urbana, IL. D. S. acknowledges support from the Foundation for Food and Agriculture Research via the new innovator program.
\end{acks}
\bibliographystyle{unsrt}
\bibliography{draft0}
\end{document}